\documentclass[preprint,showpacs,preprintnumbers,amsmath,amssymb]{revtex4}
\usepackage{graphicx}
\usepackage{bm}
\usepackage{setspace}
\begin{document}

\title{Physics of ion beam cancer therapy: a multi-scale approach}

\author{ Andrey V. Solov'yov$^{1}$\footnote{E-mail:
solovyov@fias.uni-frankfurt.de; On leave from
    A.F. Ioffe Physical-Technical Institute, 194021 St. Petersburg,
    Russia},
Eugene Surdutovich$^{1,2}$\footnote{E-mail:
surdutov@oakland.edu; Tel: +1-248-370-3409}, Emanuele Scifoni$^1$,
Igor Mishustin$^{1,3}$, and Walter Greiner$^1$}  
 
\affiliation{
$^1$Frankfurt Institute for Advanced Studies, 
Ruth-Moufang-Str. 1, 60438 Frankfurt am Main, Germany\\
$^2$Department
of Physics, Oakland University, Rochester, Michigan 48309, USA\\
$^3$Kurchatov Institute, Russian Research Center, 123182
  Moscow, Russia}
\date{\today}
\begin{abstract}
We propose a multi-scale approach to understand the physics related
to ion-beam cancer therapy. It allows the calculation of the
probability of DNA damage as a result of irradiation of tissues
with energetic ions, up to 430~MeV/u. This approach covers different
scales, starting from the large scale, defined by
the ion stopping, followed by a smaller scale, defined by secondary
electrons and radicals, and ending with the shortest scale, defined by
interactions of secondaries with the DNA. We present calculations of
the probabilities of single and double strand breaks of DNA,
suggest a way to further expand such calculations, and also
make some estimates for glial cells exposed to radiation.
\end{abstract}


\maketitle

\section{Introduction}

Ion beams are becoming more commonly used for cancer therapy as a favorable
alternative to conventional photon therapy, also known as
radiotherapy~\cite{Kraft05,Hiroshiko}.  From the physical
point of view, their advantages are
related to the fundamental difference in the linear energy transfer
(LET) by a massive  charged particle as compared with massless photons,
namely by the the presence of a Bragg peak in the depth-dose
distribution for ions. It is 
due to this peak that  the effect of irradiation on deep tissue is more
localized, thus increasing the efficiency of the treatment and reducing side
effects. In order to plan a treatment, a number of physical
parameters, such as the energy of projectiles, intensity of the beam,
time of exposure, {\em etc.}, ought to be defined. At present, their
choice is based on a set of empirical data and the experience of
personnel. Moreover, the optimization of treatment planning requires
understanding of microscopic phenomena, which take place on time
scales ranged from $10^{-22}$~s to minutes or even longer times. Many
of these processes are not sufficiently studied. Thus, a
reconstruction of the whole sequence of events and explaining,
qualitatively and 
quantitatively, the leading effects on each structural level scale
presents a formidable task not only for physics but also for
chemistry, biology, and medicine.

The ultimate goal of ion-beam therapy is to destroy the tumor by
energy deposition of the projectile resulting in the disruption of DNA and
the subsequent death of the cells~\cite{Kraft05}.  This energy deposition
is associated mainly with the ionization of the medium traversed by
the ion.  The human tissues on the average consist of 75\% water,
therefore, when appropriate, we do our calculations for liquid water.
However, we ought to consider a more complicated medium when
analyzing the ion's passage through cell nuclei. It is commonly
accepted that the secondary electrons formed in the process of
ionization are mostly responsible for DNA damage, either by directly
breaking the DNA strands, or by reacting with water molecules
producing more secondary electrons and free radicals, which can also
damage DNA. Among the DNA damage types, we emphasize single strand
breaks (SSB's) and double strand breaks (DSB's). The latter ones are
especially important because they represent irreparable damage to the
DNA, if their clustering is sufficient~\cite{Kraft05}.  Local heating
of the medium in the vicinity of ion tracks may also make the DNA more
vulnerable to damage, if not melting it. This DNA damage
mechanism has been discussed in Refs.~\cite{nimb,epjd} and deserves a
more thorough study, which is still in progress.

\begin{figure}
  \includegraphics[height=.4\textheight]{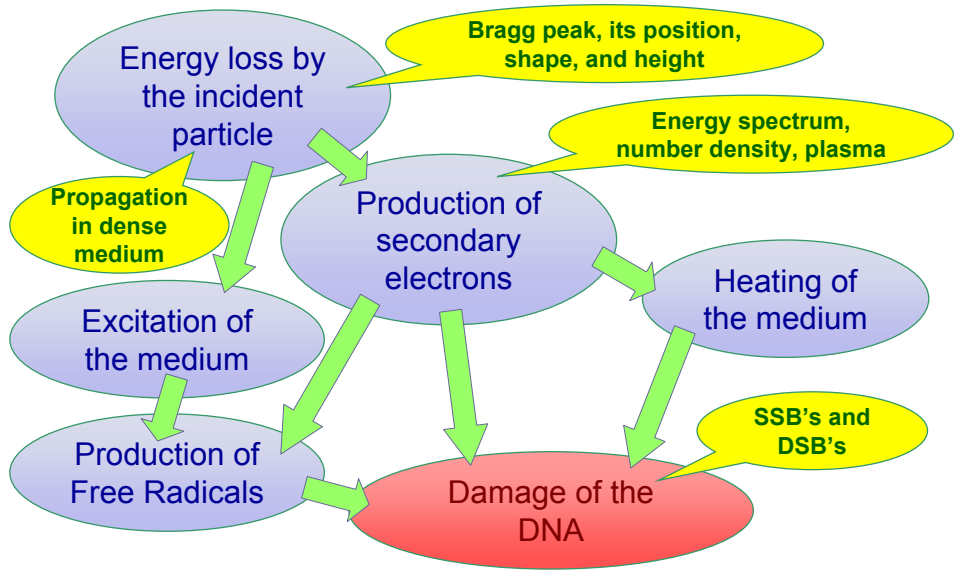}
 \caption{Schematics of the multi-scale approach.}
\label{scheme}
\end{figure}

After a fast  ion enters the tissue, many 
processes take place on different temporal and spatial scales until
tumor cells die. The goal of our approach is to analyze these
processes and identify the main physical effects which are responsible
for the success of the ion-beam therapy. It turns out that many
important aspects  should be considered in such an analysis, as is
illustrated in Fig.~\ref{scheme}. 
As can be seen from this figure, propagation and
stopping of incident ions in the tissue  represent the initial stage of the
whole scenario of ion therapy. The ion penetration depth depends
strongly  on their initial 
kinetic energy. A sharp maximum
in energy deposition close to the end of their range is called a 
Bragg peak. Many works devoted to calculation of the exact
location and shape of this peak include both deterministic and
Monte-Carlo methods, see {\em e.g.}, Ref.\cite{Pshen} and references
therein. Using the 
information about cross sections of 
atomic processes (such as ionization of water molecules) and nuclear
processes (such as
nuclear fragmentation of projectiles)  as an input, these
models give very good predictions of all characteristics of the Bragg
peak, its position, height, tail, {\em etc.} These models provide
reasonable information on the energy deposition on a mesoscopic scale
of about 0.1~mm, which is sufficient for the treatment planning. The
kinetic 
energy of ions changes from the initial energy in the range of
200-430~MeV/u down to about 50~keV/u.  
In our works~\cite{nimb,epjd}, we presented a simple
approach based on the singly differential cross
section (SDCS) of ionization of a medium.
In Ref.~\cite{nimb}, we have considered  the  ionization of water 
as a single process taking place on this scale, leaving other atomic
interactions for later consideration. In Ref.~\cite{epjd}, we included
the excitation of water molecules by projectiles. Even though no
secondary electrons 
are produced in this process, it affects the energy loss and, therefore,
the position of the Bragg peak; the excited water molecules are also prone
to possess a higher probability for dissociation, leading to free
radicals, H$\cdot$ and OH$\cdot$.  

The next scale is defined by
the secondary electrons and free radicals produced as a result of
ionization and excitation  of molecules of the medium. The maximum
energy of electrons hardly exceeds 100~eV and 
their displacement is of the order of 10-15~nm.
The main event  on this scale is the diffusion of free electrons and
radicals in the medium. They induce many chemical reactions which are
important for the DNA damage since 
they define the agents interacting with the DNA. This aspect has been
studied within several Monte Carlo models, see {\em
  e.g.}~\cite{Nikjoo06},  which use various SDCS for ion and electron 
energy loss and include effects of the
medium~\cite{liqh2o,Meesungnoen02,mfp,Pimblott3}). The propagation of electrons
and other species through the medium is simulated explicitly until
their interaction with the 
DNA. In this paper, we present another approach to describe this
stage without using Monte Carlo simulations. 
 
Interaction of electrons and radicals with DNA also
happens on a nanometer scale, and many works are devoted to study
these interactions~\cite{DNA2,DNA3,Sanche05,DNA1,DNA4,DNA5}.
In this work,
we use the experimental results of Ref.~\cite{DNA2,DNA3,Sanche05}.

The paper is organized as follows. In Section II we introduce the
various phenomena defining scales involved in the process, ion
stopping, propagation of secondary electrons and damage to
DNA. In section III, we make some estimates of DNA damage caused by
ions passing through glial cells using the results obtained in the
previous chapters and biological data.  A section of conclusions
summarizes the paper.

\section{DNA damage as multi-scale process}
\subsection{Ion stopping and production of secondary electrons}

Delta electrons play a major role in the energy
loss by projectiles and therefore determine, to a large extent, all
characteristics of the Bragg peak. Their energy spectrum, analyzed in our
previous works~\cite{nimb,epjd}, is important for the DNA
damage calculations (see below). 

Physically, the SDCS is determined by the properties of the medium,
and since we use liquid water as a substitute for biological tissue,
it is determined by the properties of water molecules and the
properties of liquid water as a continuous medium. This information
is contained in the real and imaginary parts of the electric susceptibility of
liquid water. This approach can be generalized for any real tissue if
the quantities, such as SDCS, for this medium are known.

In Refs.~\cite{nimb,epjd}, we have used for the SDCS a semi-empirical
parametrization by 
Rudd~\cite{Rudd92} and obtained the position of the Bragg
peak with a less than 3\% discrepancy as compared to Monte Carlo
simulations and experimental data~\cite{epjd}. These calculations are
not very sensitive to the exact form 
 of the SDCS, since the Linear Energy
Transfer (LET) is determined after integration over the energy of the secondary
electrons, $W$; however, the calculations of
the DNA damage may be more sensitive to the shape of the SDCS at small
energies, which for liquid water is different from
that of water vapor~\cite{Pimblott,Pimblott2,Dingfelder,Garcia}.

The SDCS is a function of the projectile's velocity  and, since the
ions are quite fast in the beginning of their trajectory, it has to be
treated relativistically. In Ref.~\cite{epjd}, this issue has been
solved by ``relativization'' of the Rudd parametrization by fitting it
to correct Bethe asymptotic behavior in the relativistic limit.

Another important issue related to SDCS is the effect of charge
transfer, due to pick-up electrons by the initially fully
stripped ions (such as $^{12}$C$^{6+}$) as they slow down in the
medium. Since the SDCS is proportional to the square of ion charge,
its reduction strongly influences such characteristics as the height
of the Bragg peak, secondary electron abundance, etc. In
Ref.~\cite{epjd}, we solved this problem by introducing an effective
charge taken from~\cite{Barkas63}. As a result, the effective charge
of the $^{12}$C$^{6+}$ near the Bragg peak is about $+3$ rather than
$+6$.

Even after the relativistic treatment of the projectile and the
introduction of effective charge, the profile of the Bragg peak
obtained in our calculations was substantially higher and narrower
than those obtained by Monte Carlo simulations or experiments. The
main reason for the discrepancy was that our calculations were
performed for a single unscattered ion, while in simulations, as well
as in experiments, the ultimate results are a combination of many ion
tracks with a significant spread in energy and position due to
multiple scattering by water molecules. After we took into account
straggling of the ions, the shape of our Bragg peak matched the shape
predicted by the Monte Carlo simulations with nuclear fragmentation
channels blocked~\cite{RADAM2}.

The nuclear fragmentation in the case of carbon ions is quite
substantial and should not be neglected. In principle, we can include
the beam attenuation due to nuclear reactions given the energy
dependent cross sections of these reactions, as we showed in
Ref.~\cite{RADAM2}. Then we would reproduce the attenuation of the ion
beam , secondary electron 
production due to different species, the spread of the Bragg peak due
to different penetration depths of different species, and the tail
following the Bragg peak due to light products such as protons and
neutrons. All these complications, however, were beyond our primary
goal of gathering the most significant effects. We leave this to
future refining calculations. We should mention that a successful
treatment of nuclear processes has been done by  the GEANT4 based Monte Carlo
simulations~\cite{Pshen}.

Thus, the ionization energy loss by ions in liquid water is the
dominating process for ion stopping and the energy spectrum of the
secondary electrons. Additional energy losses are associated with the
excitation of water molecules leading to the production of free
radicals. The SDCS defines both the longest (in
distance) scale related to the ion energy loss and provides the
initial conditions for the next scale related to the propagation of
the secondaries. 

\subsection{Propagation of  secondary electrons}

Even though the SDCS that we have used in~\cite{epjd} was not ideal,
it does give some important predictions, which
agree quite well with other calculations and measurements. Indeed, the
average energy of the secondary electrons,
\begin{eqnarray}
\left\langle W \right\rangle= \frac{1}{\sigma_{\rm T}}\int_0^\infty W
\frac{d\sigma (W,T)}{dW} dW~, 
\label{wav}
\end{eqnarray}
 in the vicinity of the Bragg peak ($T\approx 0.3$~MeV/u) is about
45~eV. This value constrains possible further processes with such
electrons. For instance, it has been shown that such electrons may excite
or ionize another water molecule, but, most likely, only once, and the
next generation of electrons is hardly capable of ionizing
water molecules~\cite{nimb,epjd}. This puts a limit on the number of
secondary electrons produced.

The secondary electrons propagate in the same medium as the ion, and
interaction with the medium is again determined by the SDCS with
electrons being projectiles. The interaction can be elastic or
inelastic, and there is a probability that it will interact with a DNA
molecule and cause damage.

The angular distribution of the secondary electrons at energies about
and below 45~eV is rather flat~\cite{angle}. Therefore, to a first
approximation, we can consider Brownian motion of secondary
electrons and use a random walk to describe their
propagation through the medium from the point of production. The
probability density to diffuse through 
a distance $r$ after $k$ steps is given by expression~\cite{Chandra}
\begin{eqnarray}
P(k, r)=\frac{1}{\left(\frac{2\pi
    kl^2}{3}\right)^{3/2}}\exp\left(-\frac{3 r^2}{2 k l^2}\right)~,
\label{rwalk}
\end{eqnarray}
where the mean free path $l$ is the average distance that
is traversed by the electron between two consecutive elastic
collisions. It is determined by the elastic SDCS for electrons 
as projectiles and we
use the results of Ref.~\cite{mfp}. Typical values for this mean free
path (for the energies of interest) are about 0.3~nm
The mean free path for inelastic
collisions $l_{in}$ is typically about 20 times longer. So,
we assume that electrons mainly experience elastic collisions and
inelastic processes are included via an attenuation factor 
\begin{eqnarray}
\epsilon(k,W)=\mathcal{N}^{-1}\exp\left(-l k/l_{in}\right)~,
\label{atten}
\end{eqnarray}
corresponding to an average distance wandered $kl$; and where
$\mathcal{N}=\int_1^\infty\epsilon
dk=\exp\left(-l/l_{in}\right)/\left(l/l_{in}\right)$, is a
normalization factor, since  
(\ref{atten}) provides the distribution over the number of steps.
Both elastic and inelastic mean free paths depend on the energy of the
wandering electron. This energy is changing gradually (with a number
of steps), and strictly
speaking, the mean free path is a function of the initial energy and
the number of steps. The energies of secondary electrons are typically
between 5 and 100~eV , and their average energy near the Bragg peak is
45~eV. The simulations of
Ref.~\cite{Meesungnoen02} suggest that the maximum range of
displacement of such  
electrons is almost independent of their initial energy. This leads us to a
model where we assume that all electrons are produced at some average
energy and diffuse with the corresponding mean free paths, $l$ and
$l_{in}$, which do not change during the diffusion.

As shown in Refs.~\cite{nimb,epjd}, the number of 
these electrons produced per segment of an ion's track, $d\zeta$,
is given by $\frac{d^2 N}{dW d\zeta}\Delta W$.  This quantity is proportional
to the SDCS of ionization by the projectiles. It depends on their
kinetic energy and, therefore, on the depth in the tissue. For the
calculations in this work we used the values from Ref.~\cite{epjd} in
the vicinity of the Bragg peak, {\em i.e.}, at an ion energy
$T=0.3$~MeV/u. In this work, we assume the
energy of electrons to be 20~eV, {\em i.e.}, 25~eV lower than the average
energy of secondary electrons, and took all of them, {\em i.e.},
integrated $\frac{d^2 N}{dW d\zeta}$ at $T=0.3$~eV from
Ref.~\cite{epjd} over $W$ from 6 to 100 eV. This gives us $\frac{d
N}{d\zeta}=8.8$~nm$^{-1}$. The reason for choosing 20-eV electrons is
an analysis of 
the behavior of mean free paths~\cite{mfp}. It turns out that if the
energy of 
secondary electrons is closer to 40~eV or higher, their inelastic mean
free path, $l_{in}$, decreases. This means that they quickly lose
energy to the value of about 20~eV, at which the ratio
$l_{in}/l=20.2$. Otherwise, it would be difficult to interpret the
results of Ref.~\cite{Meesungnoen02}.    
 
\subsection{Evaluation of DNA damage}
\label{ionpass}

DNA damage, such as a Single Strand Break, is a result of a sequence
of mutually 
independent events. First, a secondary electron with a certain kinetic
energy $W$ is produced at a certain depth $x$. Then, this electron
wanders in a surrounding medium interacting with its molecules
elastically and inelastically
gradually losing energy until it becomes bound. Depending on the electron's
energy, momentum and position, there is a chance that the electron
stumbles on a DNA molecule and damages it.
Following this scenario, we design a model for calculating the
probability of a SSB due to a passing ion.  

We consider the secondary electrons
produced by an ion traversing the medium at a certain distance from
the DNA and then let them wander toward a single convolution of the DNA. 
We represent this convolution of the DNA molecule
by a cylinder of length 3.4~nm and radius 1.1~nm (these
parameters are well established experimentally). We are interested in
calculating the number of electrons hitting a single convolution
because the DSB's are defined as simultaneous breaks of both DNA
strands located within a single convolution. 
Then, knowing the number of these electrons and their energy
distribution, we can calculate the probability of damage to this
convolution. If we can further assume a certain distribution (in
space) of other convolutions, we can then calculate the total damage to
the DNA molecule by averaging over all possible arrangements
of convolutions with respect to the ion track.

\begin{figure}
  \includegraphics[height=0.55\textheight]{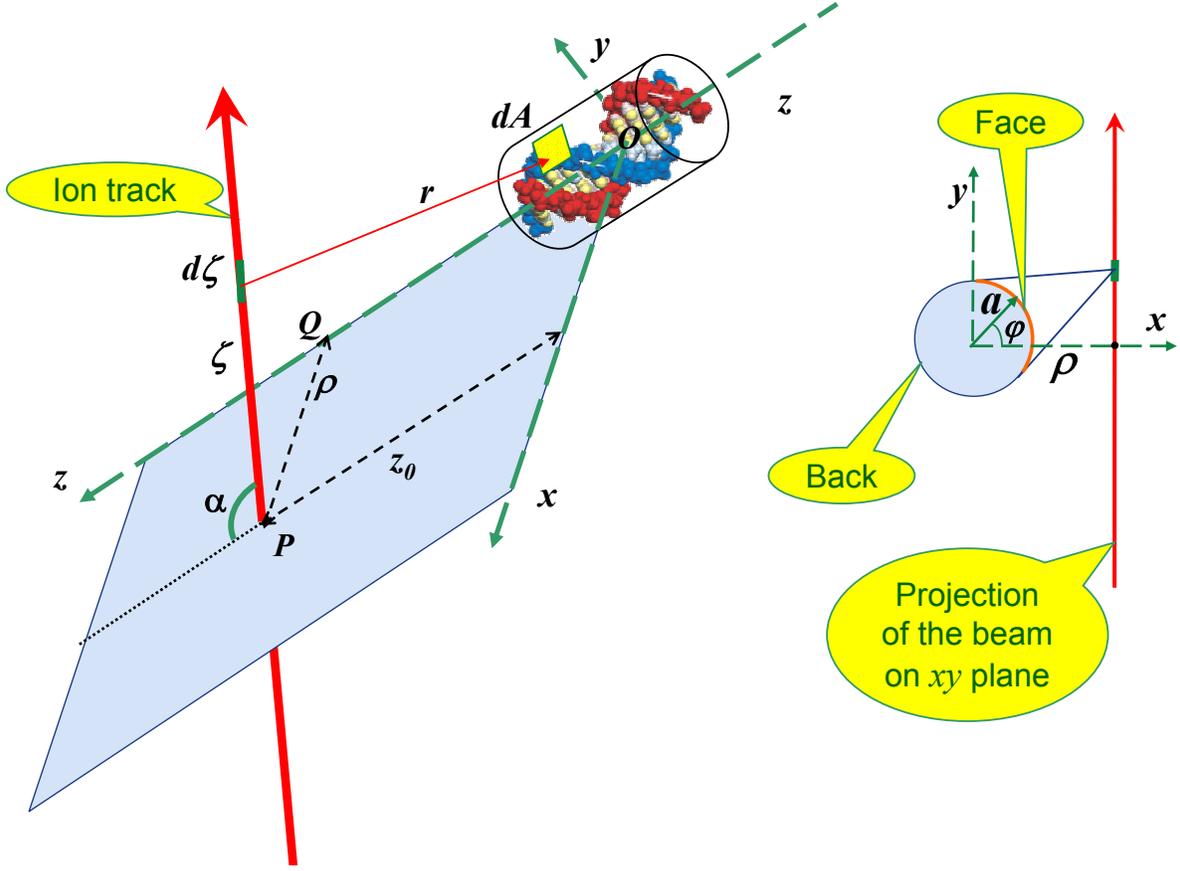} \caption{Geometry
  of the model: $z$ is the cylindrical axis of the DNA convolution and
  $x$ is chosen to be parallel to $PQ$, {\em i.e.}, the line of
  closest approach 
  between $z$ and the beam (orthogonal to both), of length $\rho$, at
  distance $z_0$ from the center of the convolution $O$. $\zeta$ is
  the coordinate of any point in the beam with respect to $P$, and
  $\alpha $ is the angle between the beam and $z$. In the right panel
  we show a projection on the $xy$ plane, where $a=1.1$ nm is the DNA
  radius and $\varphi$ is the polar angle defining a point on its
  surface.}
\label{geom}
\end{figure}

The geometrical picture of an arbitrary ion track and a chosen
convolution is schematically shown in Fig.~\ref{geom}, where
geometrical notations are also included.
From
this scheme, it stems that any distance $r$ between a point on the track
and a point on the cylinder is given by
\begin{eqnarray}
r^2=(a\cos \varphi - \rho)^2+(a\sin\varphi - \zeta \sin
\alpha)^2+(z-z_0-\zeta \cos \alpha)^2~.
\label{r2}
\end{eqnarray}
Given a number of electrons produced within a segment $d\zeta$, we
need to calculate the flux of these electrons through an 
area $d{\vec A}$ of the cylinder separated by a distance $r$ from the
segment. Assuming  Brownian diffusion of the electrons, the
probability for one electron to diffuse from the ion
track through the distance $r$ (after $k$ steps) is $P(k, r)d{\vec
r}$. 

Then the flux through a ``patch'', $d{\vec A}$, of the cylinder is given by 
\begin{eqnarray} 
d\Phi_k(r, W; W_f)=d{\vec A}\cdot D\nabla P(k, r)\frac{d^2 N}{dW d\zeta}(\zeta,
 W)\Delta W =d{\vec A}\cdot D\nabla P(k, r)\epsilon(k, W)\frac{d
N}{d\zeta}(\zeta)~,
\label{flux}
\end{eqnarray}  
where $D=kl^2/6$ is the diffusion coefficient multiplied by the
average time of wandering, and $\epsilon(k, W)$ is defined in
Eq.~(\ref{atten}).

Finally, we assume that the number of the SSB's within the DNA
cylinder is proportional to the number of electrons {\em crossing} its
surface, regardless of whether they are going into or leaving the
cylinder. This number is proportional to the integral of the absolute
value of the flux, {\em i.e.},
\begin{eqnarray}
N_{SSB}=\Gamma_{SSB}(W)\sum_k \int d\zeta d\Phi_k(r, W; W_f)~,
\label{mult}
\end{eqnarray}
where the integrations are done over the surface of cylinder and the ion
trajectory, and summation is done over the number of steps. 
The unknown
quantity $\Gamma_{SSB}(W)$, that for the 
moment we assume to be a constant ($\Gamma_{SSB}(W)=5\times10^{-4}$),
is determined 
from the experimental 
data of Refs.~\cite{DNA2,DNA3,Sanche05},
where SSB's and DSB's were induced by 0.1--30-eV electron beams.  

The calculation for the general case, as shown in Fig.~\ref{geom}, depends
on three geometrical parameters that define the position and
orientation of the DNA convolution with respect to the ion track, the
distances $\rho$ and $z_0$ and the angle $\alpha$. In order to
interpret these results, we set $z_0$ to zero, and
consider separately two limiting cases, the ``parallel'' case when
$\alpha=0$, and the ``normal'' case when $\alpha=\pi/2$.  In the
parallel case, the cylinder containing the DNA convolution is parallel
to the ion track and $\rho$ is the distance between the axis of the
cylinder and the track. In the normal case, the axis of this cylinder
is perpendicular to the ion track and when $z_0=0$, $\rho$ is again
the distance between the axis of the cylinder and the track. In the
latter case the 
beam projects along $\rho$ to the center of the cylinder.  In both
these cases, we need to set the limits for the angular integration over
$\varphi$.

Looking from any point on the ion track, there are two surfaces of the
DNA cylinder: the ``front'' or ``face'' surface and the ``back''
surface (see Fig.~\ref{geom}). In our model, if a wandering electron
hits the face or the back surface, it may cause a strand break with a
certain probability. Therefore, we simply add the probability of a SSB
due to electrons striking the back surface to that for the face
surface, regardless of the directions of their motion, leaving out the
introduction of an attenuation mechanism to account for an electron passage
``through the DNA'' for a future extension of this model.

\begin{figure}
  \includegraphics[height=.3\textheight]{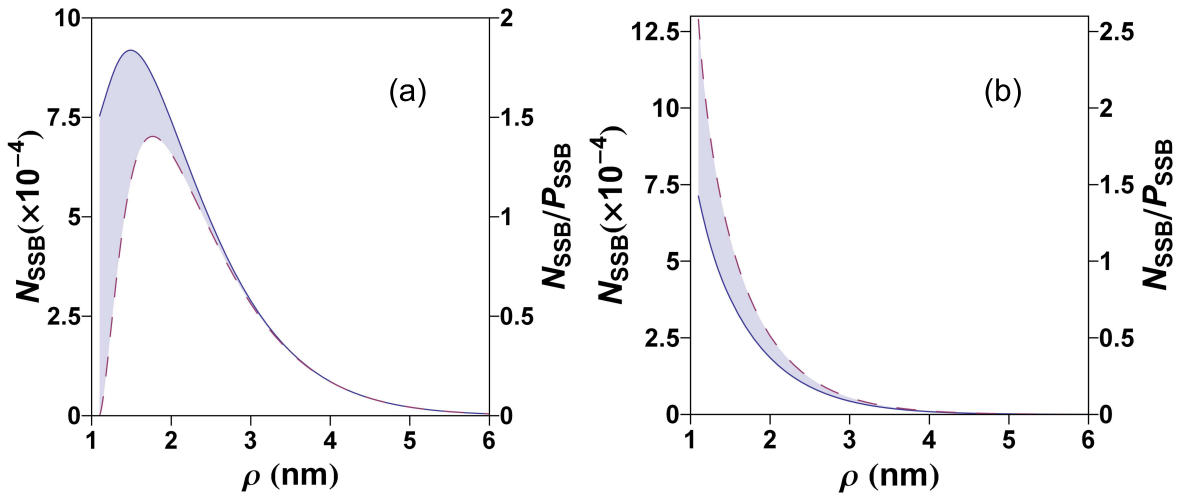}
  \caption{A comparison of numbers of SSB's for parallel (dashed line)
  and normal (solid line) configurations on the face (a) and on the
  back side (b) for 20-eV electrons.}
\label{parandnorm}
\end{figure}

\begin{figure}
  \includegraphics[height=.3\textheight]{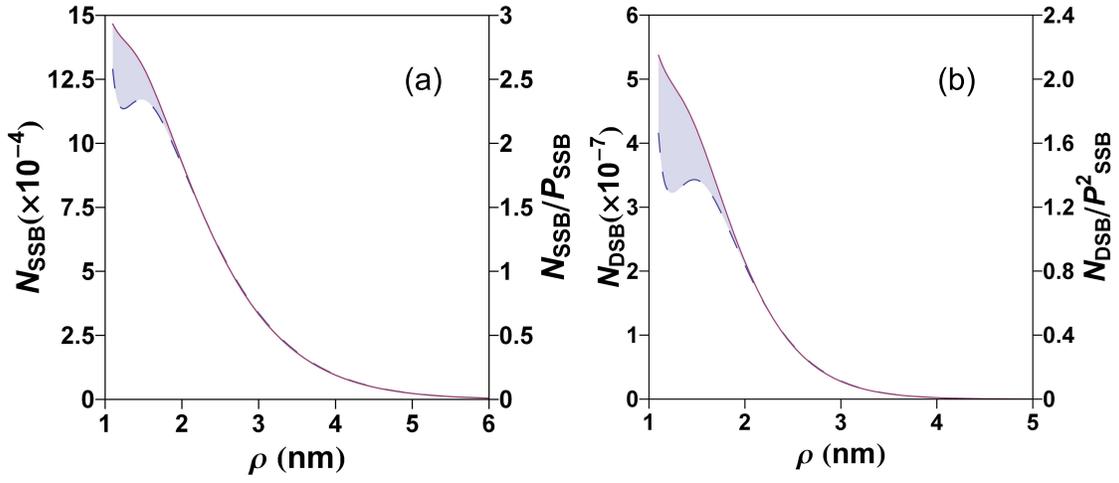} \caption{A
  comparison of dependencies of overall (due to the whole surface of
  the cylinder) SSB's (a) and DSB's due to separate electrons (b) on
  distances to the DNA convolution in the parallel (dashed line) and
  normal (solid line) cases for 20-eV electrons.}
\label{ssbdsb}
\end{figure}

\begin{figure}
  \includegraphics[height=.5\textheight]{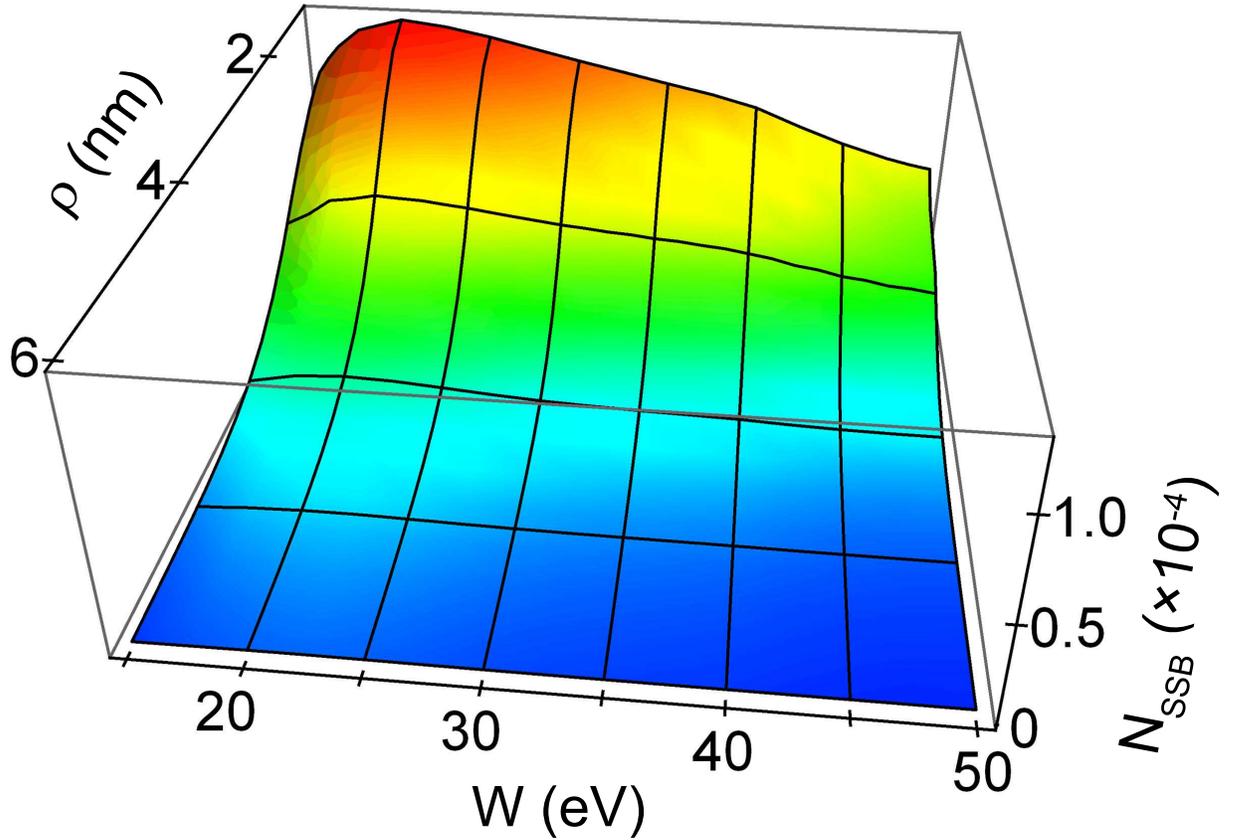} \caption{The
  number of SSB's as a function of the distance to the DNA convolution,
  $\rho$, and the energy of secondary electrons (parallel case).}
\label{ssbfig}
\end{figure}

The results of the integration are shown in Fig.~\ref{parandnorm}.
All curves show the dependence on the distance $\rho$ from the DNA to
the beam. When the distance is large enough ($>3$~nm) the normal and
parallel cases coincide for the face side as well as for the back
side; differences appear only at small distances. These differences
are significant only for back and face sides taken separately, but not
for their combination shown in Fig.~\ref{ssbdsb}a.
This means that the geometrical details of the orientation of DNA segments
with respect to the beam may not be so significant, since all
variations lie somewhere in the shadowed region in Fig.~\ref{ssbdsb}a,
{\em i.e.},
between the two curves. For a more general picture,
some average curve (lying between the two curves in Fig.~\ref{ssbdsb})
should be used, with $\rho^2$ replaced by $\rho^2+z_0^2$. The numbers
of SSB's caused 
by the secondary electrons depending on the distance $\rho$ and the
energy of the secondary electrons for the parallel case are shown in
Fig.~\ref{ssbfig}.
The decline of the number of SSB's with increasing $\rho$ is an
explicit consequence of Eqs.~(\ref{rwalk}) and
(\ref{r2}). The energy dependence is mainly caused by the dependence
of the mean free paths on the energy and the attenuation factor given
in Eq.~(\ref{atten}). This factor is heuristic and may have to be
corrected later 
when the corresponding experimental or computational data are
available.  This figure corresponds to our model, which includes
the dependence of the mean free paths on the {\em initial} energy of
secondary electrons, but assumes that the mean free paths do not
change during the 
diffusion. 

Generally, the DSB may be caused either by a single electron, if its
energy is high enough, or by two different electrons, if the electron
density is high enough. From Refs.~\cite{DNA2,DNA3,Sanche05}, it
follows that the DSB's caused by the electrons with energies higher
than about 5~eV happen in one hit, {\em i.e.}, if a particular
electron with a probability of about 0.0005 causes a SSB, the same
electron causes a DSB with a probability of about 0.2  of the
probability of the SSB (so the overall probability of a DSB is about
$10^{-4}$ rather than $10^{-7}$. This is why the analysis of the
probabilities of SSB's is so important.  If the energy of the
secondaries are high enough they give the probability of DSB's after
being divided by some factor (about 5). Therefore, Fig.~\ref{ssbfig}
also gives the shape of the dependence of DSB's on distance and energy.

At energies lower than 5~eV the situation changes; one electron 
can not produce two breaks alone. Therefore, we need to calculate the
number of DSB's caused by two different electrons. From the geometry
of a DNA convolution we infer that the probability of a DSB is
proportional to $\frac{1}{4}(N_{SSB, face}^2+N_{SSB,
back}^2)+\frac{1}{2}N_{SSB, face}N_{SSB, back}$. This factor
accounts for the probability of the occurrence of a SSB on the face for
one strand, accompanied by a SSB on the other strand occurring either
on the face or on the back, plus the probability of the inverse
event.  The numbers for the DSB's caused by different electrons in
parallel and normal cases are shown in Fig.~\ref{ssbdsb}b. Similar to
the case of SSB's, the differences due to geometry are not very
significant.  Even though the numbers of DSB's plotted in
Fig.~\ref{ssbdsb}b are many times smaller than those in
Fig.~\ref{ssbdsb}a, this effect may be significant if the density of
secondary electrons is large enough. According to our estimates in
Refs.~\cite{nimb,epjd}, the density of the secondary electrons 
produced by carbon ions at therapeutic energies in the vicinity of the
Bragg peak is by about 16 orders of magnitude higher than the electron
density in experiments of Ref.~\cite{DNA2,DNA3,Sanche05}. Therefore,
the two-electron mechanism of DSB formation may be the dominant channel.

This concludes our approach to calculations of DSB's and SSB's due to
secondary electrons produced by ions. In the following subsection
we'll analyze the next generation of secondaries produced by
electrons and free radicals produced by the ions.

\subsection{Other secondaries}

Secondary particles, which can be treated in a similar way as the
secondary electrons are OH$\cdot$
radicals. They are formed as a result of the ionization of water molecules
by an ion after dissociation of a water ion into OH$\cdot $ and
H$^+$. These radicals are formed almost at the same place as the
secondary electrons. The difference is, of course, a different
diffusion coefficient, and a different time of getting to the DNA, which
is by about 100 times longer than that for secondary electrons. Then,
the DNA damage caused by OH$\cdot$ may also be
different~\cite{sonntag,geront}. Nonetheless, if the effect produced
by OH$\cdot$ is important, this is an argument for its
inclusion into the model.  The same can be said about those free
radicals that are 
formed as a result of excitation of water molecules by ions. These
radicals (OH$\cdot$ and H$\cdot$) are also produced in the ion
track and can be treated in  a similar way as secondary electrons.

Other secondaries, such as the second generation of electrons
produced by the first generation via ionization of water, radicals
produced as a result of this process and the radicals H$\cdot$
produced via interaction of secondary electrons with water molecules
({\em e.g.}, through dissociative attachment) can be treated in the
following way. Let the interaction that produces a ``desired agent''
happen at some point ${\vec r'}$. Then the previous procedure has to
be divided into three parts: diffusion of the secondary electron from
the point of origin (the ion's trajectory) to ${\vec r'}$, an
interaction that leads to the production of the agent at ${\vec r'}$,
and the diffusion of the agent to the DNA cylinder. Then, integration
over ${\vec r'}$ has to be performed.

\section{Calculation of lesion density along the track}

We have established the way to calculate the probability of an SSB
or a DSB at a DNA convolution at a certain distance from the ion
path. Now a question is how many convolutions are there at different
distances from the ion track. If we answer this question, we can predict
the total amount of SSB's and DSB's caused by a single ion.

In order to answer this question, we have to accept a certain model for
the tissue that is being irradiated. Let us  consider glial cells in
the mediodorsal thalamic nucleus as a target. Glial cells comprise
90\% of the human brain and those of mediodorsal thalamic nucleus have
been studied experimentally~\cite{sz1}. These studies have
provided the measurements of glial cell density as well as the size of
their nuclei vital for our calculations.

Our calculations in the previous section are done for ions in the
vicinity of the Bragg peak. The effective length along the ion track
relevant for a single convolution is several nanometers, while the
width of the Bragg peak is about 1~mm. Therefore, for the first
estimate we can assume the probability of DSB's to be constant
if the distance from the ion track to the
DNA convolution is between zero and $\rho_0\approx 10$~nm and zero at
larger distances. 
Then, we can consider a cylinder of radius $\rho_0$
and length of $L=1$~mm surrounding a track of the ion traveling
through the brain tissue. According to Ref.~\cite{sz1}, the density of
glial cells is $ n_{cell}\approx 4.5\times 10^{-4}$~cells/$\mu$m$^3$, 
the characteristic cell size is  $s_c=n_{cell}^{-1/3}\approx 13 \mu$m
and the volume of a cell nucleus is $v_{nucl}\approx140~\mu$m$^3$ 
(approximate diameter of a nucleus $s_n\approx 5.2 \mu$m). Therefore, the
relative volume ``filled'' by the cell nuclei is equal to
$v_{nucl}\times n_{cell}= 0.063$, and the ion moving along this
cylinder passes through $Ln_{cell}^{1/3}\approx 75$ cells.

If glial cells are in the interphase, {\em i.e.}, the state of the
cell cycle in  which the cell exists most of its time, we can in the first
approximation assume that the DNA is uniformly distributed inside the cell
nucleus. Each DNA molecule has about $6\times 10^8$ convolutions,
therefore the average convolution density inside an interphase nucleus
is $n_{conv}=6\times 10^8/v_{nucl}=4.3\times 10^6$
convolutions/$\mu$m$^3$.

Now we have to determine the probability of a DSB, $P_{DSB}$, assuming
that they are made by single electrons of sufficient energy. From
Fig.~\ref{ssbdsb}a, we can take the value of 
$P_{SSB}=1.25\times10^{-3}$, corresponding to some average geometry, and
divide it by a factor of 5 to get a value of $0.25\times10^{-3}$
corresponding to probability of a DSB (due to the same electron).
This value is obtained for liquid water. This medium may describe
reasonably well the macroscopic quantities such as a Bragg peak
position, which rely on a high percentage of water in the tissue. Yet,
in the case we are now considering, the projectile is moving
through a cell's nucleus and the secondary electrons over this distance
are produced by ionization of its molecules, which are not only water
but also in DNA and RNA bases, sugars, {\em etc.}, in not negligible
quantities. 
 Strictly speaking, we should recalculate all cross sections
corresponding to the constituents of the cell nucleus, but for our
estimate we will take the data from Ref.~\cite{Terrissol}, which
suggest that these cross sections are about factor of 20 higher than
that for water. Therefore, considering an estimation for the ratio of
the volume occupied by these molecules in a nucleus,
$v_{chromatine}/v_{nucl }\approx$ 50\% \cite{volratio}, we can combine
these cross sections with the proper weight, obtaining
$P_{DSB}=0.25\times10^{-3}\times(20\times0.5+0.5)=2.5\times10^{-3}$.

Hence, if the ion is traveling through a cell nucleus, the
density of DSB's that it causes is equal to the $P_{DSB}\times
n_{conv}\times\pi\rho^2\approx 2.5\times10^{-3}\times 4.3\times
10^6\pi\times 0.01^2=$ 3.3 DSB/$\mu$m.  This is comparable to
the observations by Jacob {\em et al.}~\cite{Jacob} on human
fibroblasts, where, at the energy deposition of about 1~Gy
(=J/kg), the lesion density was found to be 4~DSB/$\mu$m in a nucleus.

Thus, we estimate that each ion in the vicinity of the Bragg Peak
touches 75 cells, that is $75\times\frac{s_n^2}{s_c^2}=12$
nuclei, and inside each of these nuclei, about $3.3\times s_n=17.2$
DSB's.  If a cell is not in the interphase, but rather going through
mitosis, then the DNA molecule is much more compact. Chromosomes
rather than cell nuclei become targets for the secondary
electrons. The volume of a chromosome is about 1.7~$\mu$m$^3$, which
increases the DNA convolution density to $3.5\times 10^8$
convolutions/$\mu$m$^3$, but reduces the effective volume
amenable to damage as well as the portion of the volume occupied by
molecules different by water, thus affecting the cross sections.
Then the number of DSB's per cell will be about 0.5 if
we use the same logic as above.

\section{conclusions}

Thus we presented a multi-scale approach to describe the physics
relevant to ion-beam cancer therapy. We intend to present a clear
physical picture of the events starting from an  energetic ion
entering the tissue 
 and finally leading to DNA damage as a result of this incidence. We view this
scenario as a palette of different phenomena that happen at different
time, energy, and spatial scales. From this palette, we choose the
processes that  adequately describe the leading effects and then
describe ways to include more details. We think that
calculations in this field can be made inclusively without dwelling on
a particular time or spatial scale. Our calculations are physically
motivated, time effective and can 
provide reasonable accuracy. They show that the seemingly
insurmountable complexity of the geometry of the DNA in different states
may be tackled efficiently because the geometrical differences, shown in
Fig.~\ref{ssbdsb}, are insignificant. 
We made the first estimate of the amount of DSB's caused by an ion
passing through glial cells and thus demonstrated the strength of this
approach.
We would like to encourage
experimentalists to provide data more relevant to the actual
conditions of irradiation, especially on the smallest scales involving
DNA damage. This information is vital for further tuning of our
approach by selecting and elaborating on the most important aspects of
the problem.

\begin{acknowledgments}
This work is partially supported by the European Commission within the
Network of Excellence project EXCELL and the Deutsche
Forschungsgemeinschaft; we are grateful to A.V. Korol, J.S. Payson,
I.M. Solovyova, and 
I.A. Solov'yov for multiple fruitful discussions.
\end{acknowledgments}

\end{document}